# Optical manipulation of the Rashba effect in germanium quantum wells

*Simone Rossi[1], Enrico Talamas Simola[2†], Marta Raimondo[1], Maurizio Acciarri[3], Jacopo Pedrini[1], Andrea Balocchi[4], Xavier Marie[4], Giovanni Isella[2], Fabio Pezzoli[1*]*

[1] Università degli Studi di Milano-Bicocca, Dipartimento di Scienza dei Materiali, LNESS and BiQuTe, Via R. Cozzi 55, 20125 Milan, Italy

[2] LNESS and Dipartimento di Fisica del Politecnico di Milano, Polo di Como, Via Anzani 42, I-22100 Como, Italy

[3] Università degli Studi di Milano-Bicocca, Dipartimento di Scienza dei Materiali, MIBSOLAR, Via R. Cozzi 55, 20125 Milan, Italy

[4] Université de Toulouse, INSA-CNRS-UPS, LPCNO, 135 Ave. de Rangueil, 31077 Toulouse, France

[†]Present address: Dipartimento di Scienze, Università degli Studi di Roma Tre, V.le G. Marconi 446, 00146 Rome, Italy

[*]fabio.pezzoli@unimib.it

**The Rashba effect in Ge/Si$_{0.15}$Ge$_{0.85}$ multiple quantum wells embedded in a p-i-n diode is studied through polarization and time-resolved photoluminescence. In addition to a sizeable redshift arising from the quantum-confined Stark effect, a threefold enhancement of the circular polarization degree of the direct transition is obtained by increasing the pump power over a 2kW/cm$^2$ range. This marked variation reflects an efficient modulation of the spin population and is further supported by dedicated investigations of the indirect gap transition. This study demonstrates a viable strategy to engineer the spin-orbit Hamiltonian through contactless optical excitation and opens the way towards the electro-optical manipulation of spins in quantum devices based on group-IV heterostructures.**

**Introduction**

The ability to manipulate the spin degree of freedom is of prime interest in spintronics and in the burgeoning field of quantum technologies. To this purpose, electrically induced control of the spin dynamics through spin-orbit coupling (SOC) would be practical and thus highly desirable. The spin-orbit Hamiltonian is known to give rise in solids to intriguing spin textures governing a wealth of fascinating phenomena like the spin Hall and the Edelstein effects[1–3]. SOC introduces a momentum-dependent removal of the spin degeneracy through the so-called Dresselhaus and Rashba terms of the system Hamiltonian, which arise due to bulk (BIA) and structural (SIA) inversion asymmetry, respectively. While the former occurs when nonidentical atoms compose the crystal motifs, the latter emerges when charge carriers are confined by a field-induced asymmetric potential[4,5]. The competing action of these two SOC terms leads to the development of compelling phenomena in quantum wells (QW)[6]. For instance, in GaAs/AlGaAs QWs BIA and SIA can compensate each other for all or some of the spatial directions depending on the crystallographic orientation. This drastically modifies the spin relaxation giving rise for instance to the so-called persistent spin helix states in [001] QWs[7]. Such unique phenomena often require suitable experimental configurations that leverage a precise control of the crystallographic orientation of the heterostructure and cannot be achieved along any arbitrary lattice direction.

SOC in low-dimensional structures based on group IV semiconductors inherently lacks the BIA term being suppressed by the centrosymmetric crystal structure. In these heterostructures the SOC Hamiltonian can be more easily engineered than in the III-V counterparts by acting directly on the SIA through an external electrical perturbation. In this context, Ge/SiGe quantum wells (QW) are particularly interesting because they offer the unmatched manufacturing capabilities of contemporary semiconductor industry[8], alongside various advantageous attributes, such as a favorable electron and hole confinement within the QW[9,10], highly-tunable polarized emission[11], strong g-factor anisotropy [12,13] and long spin lifetimes[8,14,15].

Despite such notable properties, the investigation of the Rashba physics in Ge QWs is still at its infancy, particularly for electron spins. Surprisingly, even central information like the impact of an electric field on the radiative recombination presently remains unexplored. To fill this gap, we leverage polarization- and time-resolved photoluminescence (PL) to address the spin-dependent properties of electronic devices hosting Ge QWs and $Si_{0.15}Ge_{0.85}$ barriers. Specifically, we utilize the characteristic type-I band alignment of the Ge/SiGe heterointerface and the built-in structural asymmetry of the p-i-n junction to explore the emergence of the Rashba field.

The steady state measurement of the circular polarization degree of the ultrafast direct gap recombination demonstrates that the inherent asymmetry of the diode induces splitting and orbital mixing of the electronic spin states of the QWs. This eventually modifies the selection rules regulating vertical transitions and tends to wash out the expectation value of the spin operator resulting from the process of optical spin orientation. Our findings demonstrate nonetheless that the photogeneration of long-lived electrons at the indirect gap piles up charges within the QWs. This effect is adjusted by the optical pump that partly reduces the built-in potential gradient, hence the Rashba field, reestablishing a sizeable electron spin polarization. The development of novel control strategies for SOC can further stimulate theoretical and experimental

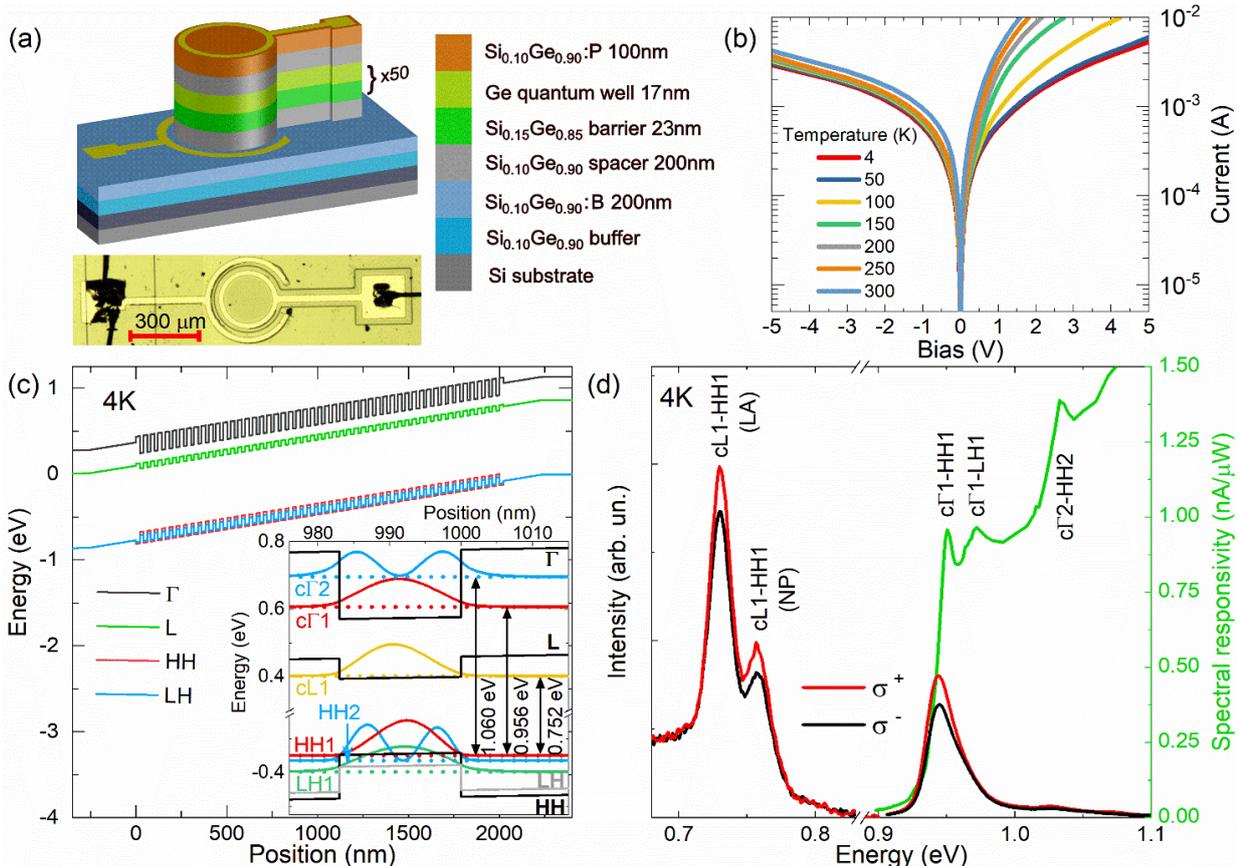

Figure 1 (a) Schematics of the epitaxial heterostructure (not to scale). An optical image of a 300 µm device is shown below. (b) Dark I-V characteristics measured at different temperatures. (c) Low temperature band edges of the main electronic states simulated with an 8 bands k·p method. The inset shows the first few levels with corresponding arrows that indicate the main transitions and their calculated energies. (d) PL of the device at 4K. The laser source is right-handed circularly polarized (σ+) and has an excitation energy of 1.165 eV and a power density of 3.40 kW/cm². The right-handed (σ+) and left-handed (σ-) circularly polarized PL component is reported as red and black lines, respectively. The PL spectrum is obtained by collating data from a linear array (>0.83 eV) and a single channel (In,Ga)As detector (<0.83 eV). The spectral responsivity of the p-i-n diode at 4K is shown as a green line.

investigations of the spin dynamics in group IV epitaxial architectures beyond silicon[8,16,17] and provides creative solutions towards the implementation of future spin-optronic concepts[18]. The investigation of Rashba SOC is also relevant in the context of spin-to-charge interconversion, leading to the exploration of spin-related phenomena that can decisively enrich the competitive potential of Ge-based systems in the realm of semiconductor spintronics and in forthcoming applications of quantum technologies.

Finally, our experimental results also provide a deeper understanding of the physics of Ge-based low-dimensional structures as we demonstrate that the electric field can deeply influence the electron-phonon interaction and suppress the parasitic luminescence of defects.

**Results and discussion**

Figure 1a shows the layout of the Ge QW p-i-n vertical diode along with an optical image of a 300 µm mesa (see method for a detailed description of the device). Fig. 1b summarizes typical I-V characteristics measured under dark conditions. As the temperature drops, the diode loses some of its rectifying behavior and the I-V curve becomes more symmetric, i.e., with the direct and reverse regimes being similar to each other. We recall that reverse current is affected by the thermal energy of the carriers, therefore the small reduction observed at negative bias can be ascribed to the diminishing probability of generation of the electron-hole pairs. In the forward regime, the pronounced decrease of the electrical current can be attributed to a reduction of the thermally activated diffusion of majority carriers above the built-in potential barrier and a partial freeze-out at cryogenic temperatures.

The whole p-i-n structure is simulated at the center of the Brillouin zone ($\Gamma$ point) through an 8-band k·p method [19] and at the zone edge (*L* point) with an effective mass approximation. Figure 1c shows the energy levels of a QW in absence of any external bias (a calculation of the biased device can be found in the Supporting Information). The noticeable bending of the band edges along the growth direction and the shift of the electron and hole probability densities in opposite directions (figure SI0) unveils the built-in voltage generated by the asymmetric doping and the emergence of the quantum confined stark effect (QCSE). The states that are responsible for the experimentally accessible optical transitions are shown along with the calculated transition energies at 4K. Note that excitonic effects are neglected. The simulation of the entire structure, comprising 50 QWs, is also shown in Fig. 1c.

Figure 1d shows the PL spectrum obtained at zero external bias by merging measurements using two different photodetectors. In this way, we maximize the signal-to-noise ratio over a broad spectral region and gather simultaneous access to radiative events occurring through both indirect and direct gaps. Besides being spectrally well-resolved, these two recombination channels are suitably characterized by very different lifetimes, namely tens of ns and hundreds of fs for the indirect and direct gap, respectively. We can therefore leverage their PL signal to gather in a single shot information on the charge and spin dynamics occurring at different time scales.

Two main features can be observed in the PL spectrum shown in Fig. 1d. At high energy, the peak at 0.945 eV can be attributed to the c$\Gamma$1 – HH1 recombination through the Ge/SiGe QWs, whereas in the low energy range the peak at 0.757 eV is associated to the no-phonon (NP) cL1 – HH1 transition and is accompanied by a replica redshifted by about 27 meV due to longitudinal acoustic (LA) phonons [20,21]. The low-energy tail of the indirect transition stems from the contribution of the broad lower energy emission of defects, chiefly dislocations[20]. Some additional weaker features occur on the high energy side of the c$\Gamma$1 – HH1 peak. The k·p calculations suggest that the kink at 1.027 eV can be attributed to the recombination involving c$\Gamma$2 and

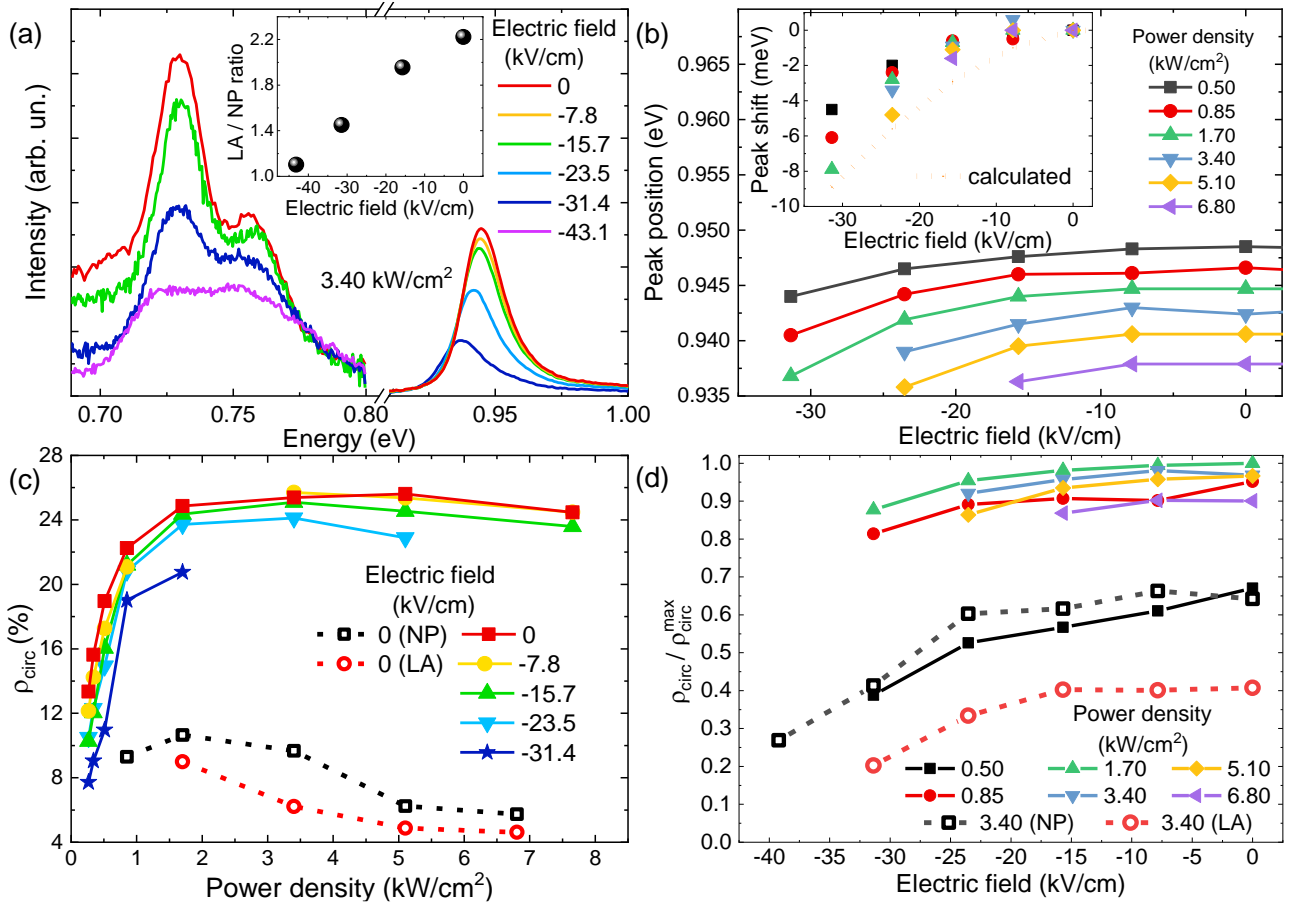

Figure 2 (a) Right-handed (σ+) circularly polarized spectra under a 3.40 kW/cm² continuous-wave illumination but at a varying electric field. (b) Direct gap PL peak positions as a function of the electric field. The inset reports the shift of the cΓ1 – HH1 peak with respect to the value extracted at 0 kV/cm for each illumination condition, while the orange dashed line represents the k·p results. (c) Degree of circular polarization for different illumination conditions. Full marks correspond to the direct transition, while open marks are about the zero phonon (NP, black) transition and the indirect transition mediated by longitudinal acoustic phonons (LA, red). (d) Degree of circular polarization as a function of the external electric field at various illumination conditions. Full marks are for the direct transition, whereas open marks represent the indirect (NP and LA) transitions (for our samples 1 V ~ 3.9 kV/cm). All the graphs are derived from low temperature (4 K) measurements.

HH2 energy levels. The identification of the PL spectral feature related to direct-gap of the PL spectra is further confirmed by the spectral responsivity also shown in Figure 1d. The small shift (~7 meV) of all the photocurrent structures compared to the associated PL counterparts evidences the so-called Stokes shift arising because of the inherent fluctuations of the QW thickness[20,22], or strain and alloy composition in the barriers. These photocurrent and PL measurements are performed at different excitaiton consitions, therefore the thermal shrinkage of the bandgap due to the relatively higher power used for the PL likely affects the Stoke shift.

In the following, we investigate the optical response of the device under the presence of an external electric bias. Figure 2a focuses on the PL at 4K and at a fixed excitation power density of 3.40 kW/cm². A field-induced monotonous reduction in the cΓ1 – HH1 emission intensity accompanies a redshift of the peak. This behavior is the experimental hallmark of the QCSE [23–25].

When the electric field varies from 0 to -31.4 kV/cm (-8V) the resulting energy shift is about 6 meV. This finding compares well with the k·p simulation (see the inset of Figure 2b) and is in line with previous literature reports on similar heterostructures[9,26,27]. Such a quantity remains almost constant at different excitation conditions, although an offset appears, drifting rigidly the QCSE behavior towards lower energies as the excitation power increases (Figure 2b). An effect that mainly stems from the bandgap shrinkage originating from a laser-induced heating.

Even though Ge/SiGe QWs host a notable type I band alignment, the fundamental nature of the energy gap remains indirect. The perturbation caused by the Stark effect on the carriers dwelling at the *L* valleys poses interesting and fundamental questions on how the electric field couples to phonon mediated transitions: a process that presently remains poorly understood from an experimental and a theoretical standpoint. While the redshift is less resolved than for the cΓ1 – HH1 peak (see Fig. SI3), here we also observe that by applying an electric field, both the cL1 – HH1 (LA) and cL1 – HH1 (NP) transitions manifest an intensity decrease as expected because of the QCSE (see Fig. 2a). Interestingly, also the defects tail (signal below 0.71 eV) loses spectral weight, which anticipates a marked reduction of the carrier recombination rate from the defects states. The most striking effect, however, is that the field-induced suppression affects the LA replica more than the NP emission. This further suggests that the bias has a stronger impact on the momentum-conserving transitions involving phonons (LA) rather than structural imperfections (NP). All these findings do not find correspondence in the well-established literature of QWs based on III-V compounds and point towards the need of dedicated studies to unleash the physics of electron-phonon interactions under QCSE perturbation. An alternative explanation for the observed phenomena can rely on the enhancement of the NP transition due to shift of the electron wavefunction towards the interfaces and the concomitant depopulation of L-valley electrons due to tunnelling processes, the latter being favoured by the shallow confining potential at the zone edge. Such a possibility, however, seems unlikely because of the well-followed QCSE-like behaviour and, as shown later, because of the absence of a significant influence of the external field on the carrier lifetime measured at the NP transition.

The extensive characterization of the optical properties of the device so far discussed enables us to further implement an all-optical investigation of the spin-dependent properties under symmetry breaking in Ge/Si$_{0.15}$Ge$_{0.85}$ QWs. The co- (σ$^+$) and counter-circular (σ$^-$) polarized PL spectra, obtained for a right-handed circularly polarized excitation, are reported in Fig. 1d as red and black lines, respectively. The polarization-resolved PL components demonstrate a considerable intensity imbalance by virtue of the net spin population generated through optical spin orientation. We therefore concentrate on the resulting degree of circular polarization of the PL, ρ$_{circ}$ (see methods), and begin the discussion with the direct gap transition when the illumination power density is varied under continuous-wave (CW) excitation.

Figure 2c demonstrates that at zero external bias (red squares) the increment of the pump power up to 2 kW/cm$^2$ suddenly leads to a rapid upsurge of ρ$_{circ}$. Since motional narrowing characteristic of BIA can be ruled out due to the centrosymmetric crystal structure of Ge, the likely culprit for this unexpected behavior can be the following. Upon illumination and in absence of electric fields, the very large majority of electrons photogenerated at the direct gap scatter out of the Γ valley because of the intense electron-phonon coupling. The resulting lifetime pertaining to electrons at the zone center turns out to be vanishingly small ($\tau_\Gamma \sim$ hundreds of fs[28]) with respect to the electron spin relaxation time ($T_1 \sim$ few μs[8]). If we now recall that under CW excitation $\rho_{circ} \propto P_0/\left(1 + \frac{\tau_\Gamma}{T_1}\right)$, where *P$_0$* is the average electron spin at the instant of photocreation induced by a hν=1.165 eV photon flux, it becomes evident that the ultrafast electron dynamics at Γ implies that the polarization of the direct gap transition is almost insensitive to the Rashba-induced modifications of $T_1$. As a result, changes of *P$_0$*, if any, seamlessly map out onto variations of ρ$_{circ}$. In other words, the polarization of cΓ1 - HH1 can be thereupon leveraged as a convenient tool to assess the impact of SIA on the spin-dependent selection rules of the optical transitions. This is in stark contrast to conventional direct gap materials, in which $T_1$ and *P$_0$* cannot be easily disentangled through steady state spectroscopy. The dependence of $\rho_{circ}$ on the pump power shown in Figure 2c further discloses extraordinary information. The value of $\rho_{circ}$ obtained in the unbiased device at the lowest power density is very small as compared to similar MQWs not embedded in a p-i-n junction, e.g., Ref. [14]. This readily manifests that the built-in bias of the diode structure causes a mixing of the spin-orbitals of the states involved in the optical transitions. The asymmetric doping of the device introduces a finite Rashba SOC, whose maximum magnitude establishes at dark condition and is given by the gradient of the static potential along the growth direction. The associated spin and momentum dependent Rashba terms in the effective Hamiltonian thereby yield a marked suppression of the electron spin polarization $P_0$ compared to a very same but symmetric Ge/SiGe QWs, in which the Rashba effect would be completely absent. The minute $\rho_{circ}$ that we observed in our asymmetric QWs in the low excitation power regime therefore supports the field-induced spin-mixing phenomena, which was originally predicted in externally biased Ge/SiGe QWs by Virgilio and Grosso[29].

The intriguing and unique property further disclosed by our sample is that even in absence of external electric fields we can enlarge the spin polarization and hence $\rho_{circ}$ by solely acting on the pump power. As mentioned before, photocreated electrons predominantly leave the zone center and accumulate at the fundamental conduction band minima located at the $L$ valleys of the Brillouin zone. This mechanism is favored by the long carrier lifetime[21] characteristics of the indirect gap and can eventually lead to a damping of the built-in field of the device. The effectiveness of this process is regulated, upon steady-state illumination, by the density of the accumulated photocarriers, that is, by the laser fluency. The most notable consequence of such pump-induced charge accumulation is the screening of the Rashba field experienced by the photogenerated electrons that remain at the $\Gamma$-valley. This mechanism accounts for the power-dependent value of the circular polarization of the direct c$\Gamma$1 - HH1 peak and is consistent with the unusual increase up to a maximum value of about 26% [14]. The data summarized in Fig. 2c and d further suggest that photo-injection is indeed the dominant factor in determining $\rho_{circ}$. This optical effect demonstrates to be very robust. In comparison, the effect of the electric field is relatively weaker (see also Supporting Information) and $\rho_{circ}$ decreases when an external bias is applied because of the enlarged amount of poorly polarized electrons recombining from the depletion region. All these findings clarify that photogeneration of carriers in asymmetric Ge/SiGe QWs can be used to dynamically control spin and orbital mixing of the electron wavefunctions. More specifically, optical excitation can be used to counteract the built-in Rashba field, providing an unmatched and effective means to optically engineer the system Hamiltonian.

In the high optical injection regime, that is, above 2kW/cm$^2$ in Fig. 2c, a very smooth decrease of $\rho_{circ}$ can be finally seen. This subtle behavior can occur because of concomitant effects, e.g., local temperature raise and, possibly, Auger-assisted excitation of unpolarized $L$-valley electrons[30]. The data in Fig. 2c manifest a similar behavior in the high-power regime also for the indirect transitions, thus suggesting that spin-relaxation due to electron-hole exchange interaction can also occur[8]. It should be noted that an unpolarized defect tail extends into the spectral region of the indirect peaks (see Figure 1d). This contribution has been removed (see Supporting Information) to single out the genuine $\rho_{circ}$ pertaining to the indirect recombination. The low polarization and signal-to-noise ratio of the zero-phonon transition and its LA replica however impede the determination of $\rho_{circ}$ in the low power regime.

Although a time-domain investigation of the direct recombination is not readily accessible through PL, we can gather informative access to the spin and carrier dynamics occurring in asymmetric Ge/SiGe QWs by focusing hereafter on the relatively long carrier lifetime of the cL1 – HH1 transition. Polarization-resolved decay curves of the NP peak measured at 4K and at different excitation densities are reported in Figure 3a (see the Supporting Information for a comparison with the excitation pulse). For each curve, two dominant

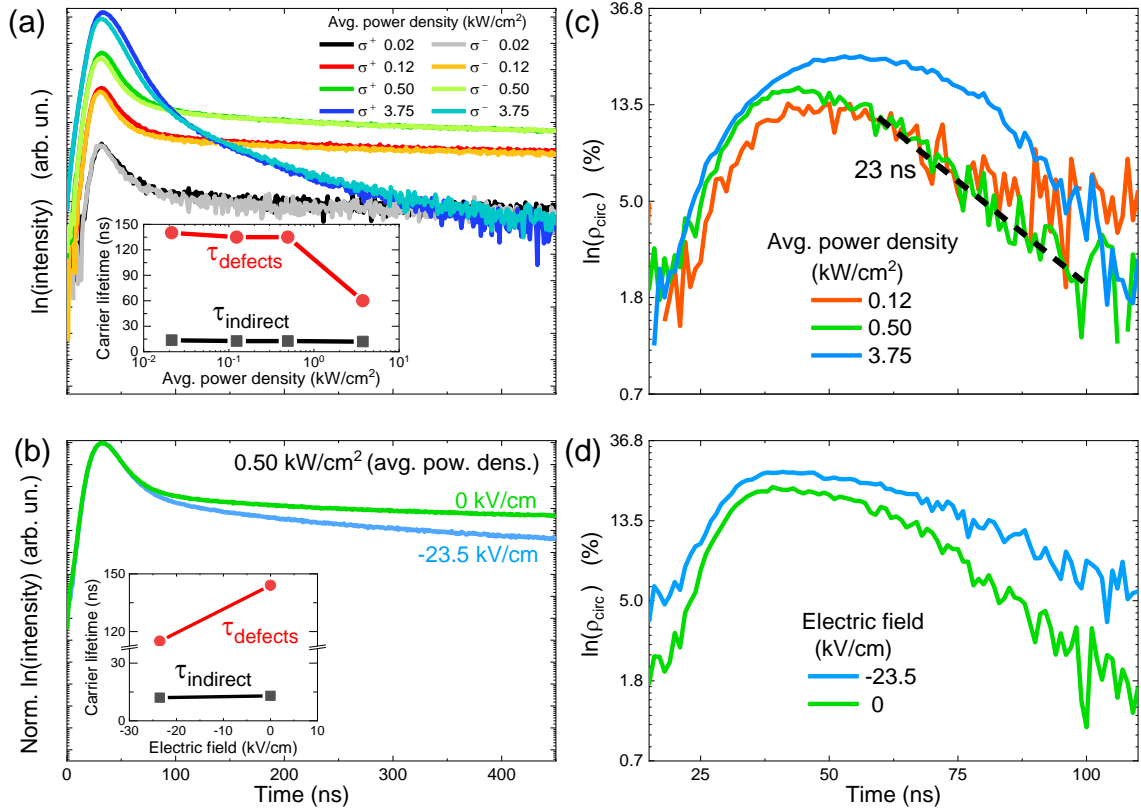

Figure 3 (a) Co- and counter-polarized decay curves measured at 0.758 eV with no applied external bias. Different power densities of the excitation are reported. The inset shows the effective carrier lifetime extracted from indirect and defects transitions by using a double decay exponential fit performed on the average of right (σ+) and left (σ-) circularly polarized curves for each illumination condition. (b) Co-polarized decay curve for different fields applied under an illumination of 0.50 kW/cm². (c) Decay curves of $\rho_{circ}$ extracted for each power density. The lowest excitation is not reported for clarity because no net polarization is found. A tentative linear fit for the 0.50 kW/cm² case is reported indicating a spin relaxation time of 23 ns. (d) Decay curves of $\rho_{circ}$ obtained for various values of the external electric field. All the measurements are performed at 4K.

components can be observed. At first, the decay is characterized by a short effective lifetime ($\tau_{indirect}$) of approximately 14 ns. A comparison with the literature data enables us to ascribe this transient feature to the recombination processes occurring in the Ge/SiGe QWs through the indirect gap[21]. An additional mono-exponential decay intriguingly emerges after ~40 ns from the beginning of the dynamics. This component can be attributed to the slower recombination from the tail of the defect states, being characterized by a long lifetime ($\tau_{defects}$) of approximately 150 ns. The most striking effect of the pump power is the marked shortening of $\tau_{defects}$, as elucidated in Fig. 3a (inset) by the compelling suppression of the defect-related tail when the excitation power level is increased. Conversely, no sizeable change of $\tau_{indirect}$ can be observed (see the inset of Fig. 3a). We can speculate that this unexpected acceleration of the kinetics of the carriers trapped at the defect sites can be rationalized within the framework of the dislocation barrier model[31]. Changes in the electrostatic potential between the charged defects and the encapsulating matrix can be either caused by an external electric field or by the charge density redistribution due to carrier accumulation upon illumination. Such a mechanism can modify the electrostatic environment, i.e., the local confining potential in the neighboring of the defect site, and can inhibit its function as a recombination pathway. We point out that the present findings do not allow us to discriminate whether such phenomenon can be due to a reduced capture cross-section of the defects rather than the activation of non-radiative Auger recombination following a rapid filling of the density of states of the traps. We can nevertheless find reassurance of the proposed physical picture from the CW data (Fig. 2a) and from the study of the PL decay when an electric field between the contact leads is established. Fig. 3b shows indeed that also in this case, $\tau_{defects}$ decreases upon the application of an external bias, consistently with a Shockley-Read-Hall model, whereas $\tau_{indirect}$ remains unaffected.

The PL data in the time domain shown in Fig. 3a directly offer a crucial information on the spin-dependent phenomena developing in the asymmetric Ge/SiGe QWs. The increase of the optical power density opens indeed a gap between the two helicity-resolved decay curves, indicating the enrichment of the out-of-equilibrium electron spin population due to a more effective optical orientation process. This is seemingly reflected in the transient of the polarization degree as derived from the PL decay curves and summarized in Fig. 3c. It should be noted that the lowest excitation has not been reported because of the negligible polarization. Remarkably, the maximum of $\rho_{circ}$ increases from ~0 to 22 %. We expect the contribution of unpolarized photons emitted through the defects to be minor at the initial phase of the decay dynamics and not to significantly affect the polarization maxima considered here. It is illuminating to notice that such power-induced effect is in full agreement with the modification of $P_0$ previously inferred from the steady-state measurements of the cΓ1 – HH1 polarization (Fig. 2c). This further supports our analysis and makes a stronger case for the suggested optical control of the effective Hamiltonian of asymmetric Ge/SiGe QWs, thus opening new perspectives for spin control in future semiconductor spintronics.

Since the polarization and time-resolved analysis of the NP cL1 – HH1 provides us with direct access to spin kinetics, we can notably explore the role of SIA on $T_1$. The Rashba splitting can indeed cause k-dependent precession of the electron spin, eventually accelerating spin loss, i.e., shortening $T_1$. The previous data obtained under CW excitation let us expect that the optical pump holds the potential for counteracting the Rashba-driven spin relaxation, thus providing an unconventional contactless strategy, alternative to typical electric fields, for manipulating the spin dynamics.

Even though the transient curves of $\rho_{circ}$ consistently unveil a non-trivial decay, possibly due to the subtle contribution from the defect states, the data summarized in Fig. 3c indicate a flattening of the curves during the initial phase of the dynamics, i.e., the one dominated by interband recombination. As a reference, a $T_1$ of about 23 ns is extracted at zero bias under an excitation power density of 0.50 kW/cm$^2$ (dashed line associated to the green curve of Fig. 3c). Figure 3c points towards a possible lengthening of the spin relaxation with the excitation power that would be consistent with the illumination-induced screening mechanism discussed above. To further consolidate this argument, we can observe in Fig. 3d that again the gross features in terms of overall polarization and modifications of the initial decay can be possibly observed also when an external electric field is applied at a constant pump power.

 Conclusions

We have demonstrated a practical implementation of the Rashba effect in group IV heterostructures using a p-i-n diode with 50 Ge/Si$_{0.15}$Ge$_{0.85}$ QWs embedded in the intrinsic region. A successful optical spin orientation is achieved showing a different emission intensity of the two circular polarizations for both the direct and the indirect emissions. Under the application of an external electric field, we observe the quantum-confined Stark effect at different illumination conditions.

The polarization-resolved photoluminescence analysis unveils a tunable degree of circular polarization with the pump power density that can be ascribed to the electrical screening of the Rashba field induced by the accumulation of the photogenerated charges. Indeed, $\rho_{circ}$ stretches between approximately 7 and 26%. Our investigation is further supported by time-resolved analysis of the indirect emission. The results also show a non-trivial kinetics due to contribution arising from defects. The indirect transition remains unperturbed by both the illumination and the electric field with an extracted $\tau_{indirect}$ of 14 ns, while there exists a suppression of the defect contribution leading to a reduction of their associated lifetime.

Future studies should be directed to address fundamental phenomena such as the electric field perturbation of the electron-phonon interaction, and to deepen our understanding of the origin of the SOC spin splitting of conduction band states at the L-valley. In fact, it remains to clarify whether the Rashba field mimics the intrinsic spin lifetime of Ge and originates from electric field in the remote bands rather than in the valence band[32]. These results constitute an all-optical study of the Rashba physics in group IV semiconductors, leading the way to future investigations of electrical-optical manipulation of spins in quantum technologies based on spin-photon interfaces.

**Methods**

A Ge/Si$_{0.15}$Ge$_{0.85}$ heterostructure was grown by low-energy plasma enhanced chemical vapor deposition[33]. A graded virtual substrate was formed starting from a (001)-oriented Si wafer by linearly increasing the Ge molar fraction at a rate of about 7%/μm until a final Si$_{0.1}$Ge$_{0.9}$ alloy was reached. The graded structure was capped by 2 μm of a Si$_{0.1}$Ge$_{0.9}$ buffer layer. The p-i-n active region was subsequently grown as follows. At first, 200 nm of Si$_{0.1}$Ge$_{0.9}$:B (4×10$^{19}$ cm$^{-3}$) are deposited prior to 200 nm of a not-intentionally doped Si$_{0.1}$Ge$_{0.9}$ spacer. Then 50 QWs of pure Ge with a thickness of 17 nm were sandwiched between 23 nm thick Si$_{0.15}$Ge$_{0.85}$ barriers to form the core of the intrinsic region. Finally, 250 nm of a not-intentionally doped Si$_{0.1}$Ge$_{0.9}$ spacer is grown, followed by 100 nm cap made of a Si$_{0.1}$Ge$_{0.9}$:P (2×10$^{19}$ cm$^{-3}$) layer. Dry chemical etching exploiting 21 Bosch cycles was applied to define circularly shaped mesas having ~2.55 μm height and diameters ranging from 200 to 500 μm. Ohmic contacts were obtained by depositing and patterning ring-shaped Ti/Al layers onto the doped regions. No annealing treatment was performed.

PL spectra were measured at 4K under the excitation of a right-handed circularly polarized Nd:YVO$_4$ laser at 1064 nm (1.165 eV). The laser spot diameter on the device was approximately 50 μm. Continuous-wave and Q-switched pulsed lasers were employed alike. In the latter case, the pulse duration was of about 15 ns and the repetition rate of 10 kHz. The PL was sent through a linear polarizer and a quarter-waveplate. The degree of circular polarization, which is a measure of the different intensity between the right- ($I^+$) and left- ($I^-$) handed circularly polarized components of the PL, was retrieved from time-resolved data as $\rho_{circ}(t) = \frac{I^+(t) - I^-(t)}{I^+(t) + I^-(t)}$, or determined through CW experiments by performing a full Stokes analysis of the PL (see Refs. [34,35] for details). The measurement of the PL intensity was conducted by coupling a monochromator to one of the following detectors: a photomultiplier and a linear array both possessing the long wavelength cutoff at ~0.75 eV, and a single channel (In,Ga)As photodiode with a cutoff at 0.52 eV. The latter required a phase-sensitive detection. Electrical bias was applied to the device through a Keithley source measure unit. Finally, photocurrent measurements were carried out using a lock in amplifier and a supercontinuum source to cover the spectral range from 0.855 to 1.078 eV.

**Acknowledgements**

The authors acknowledge T. Galliani, E. Radice and N. Radice for their technical support on optical spectroscopy. J.P. acknowledges financial support from FSE REACT-EU (grant 2021-RTDAPON-144).